\documentclass[12pt]{article}\usepackage[hyperfootnotes=false]{hyperref}
   \usepackage{epsfig}
      \usepackage{amsmath}
      \usepackage{amssymb}
  \usepackage{graphicx}
  \newlength{\abstractwidth}
  \renewcommand{\thefootnote}{\fnsymbol{footnote}}
  \renewcommand{\thanks}[1]{\footnote{#1}} 
  \newcommand{\starttext}{
  \setcounter{footnote}{0}
  \renewcommand{\thefootnote}{\arabic{footnote}}}
  \renewcommand{\theequation}{\thesection.\arabic{equation}}
  \newcommand{\be}{\begin{equation}}
  \newcommand{\bea}{\begin{eqnarray}}
  \newcommand{\eea}{\end{eqnarray}}
  \newcommand{\beq}{\begin{equation}}
  \newcommand{\ee}{\end{equation}}
  \newcommand{\eeq}{\end{equation}}

  \def\ba{\begin{eqnarray}}
  \def\ea{\end{eqnarray}}

  \def\12{{1 \over 2}}
  \def\eq{&=&}

 \def\simleq{\; \raise0.3ex\hbox{$<$\kern-0.75em
      \raise-1.1ex\hbox{$\sim$}}\; }
 \def\simgeq{\; \raise0.3ex\hbox{$>$\kern-0.75em
      \raise-1.1ex\hbox{$\sim$}}\; }

\def\O2{\Omega_2}

\def\bi{\begin{itemize}}
  \def\ei{\end{itemize}}

\def\S{Schwarzschild}
\def\sc{\setcounter{equation}{0}}
\def\pb{pull-back---push-forward}

\def\r{${\cal{R}}$}
\def\z{${\cal{Z}}$}
\def\h{${\cal{H}}$}
\def\pb{pull-back---push-forward}
\def\w{$ \omega $}
\def\W{$\Omega$}
\def\W'{$\Omega$}

\def\V{\Omega}
\def\V'{\Omega}
\def\dof{degrees of freedom}

\begin{document}
  \renewcommand{\theequation}{\thesection.\arabic{equation}}

\begin{titlepage}
  \rightline{}
  \bigskip

  \bigskip\bigskip\bigskip\bigskip

    \centerline{\Large \bf { }}
    \bigskip

    \centerline{\Large \bf { The Transfer of Entanglement:  }}

    \bigskip

    \centerline{\Large \bf { The Case for Firewalls  }}

    \bigskip

  \bigskip \bigskip

  \bigskip\bigskip
  \bigskip\bigskip

  \begin{center}
  {{ Leonard Susskind}}
  \bigskip

\bigskip
Stanford Institute for Theoretical Physics and  Department of Physics, Stanford University\\
Stanford, CA 94305-4060, USA \\

\vspace{2cm}
  \end{center}

  \bigskip\bigskip

 \bigskip\bigskip
  \begin{abstract}


Black hole complementarity requires that the interior of a black hole be represented by the same degrees of freedom that describe its exterior. Entanglement plays a crucial role in the reconstruction of the interior degrees of freedom. This connection is manifest in ``two-sided" eternal black holes. But for real black holes which are  formed from collapse there are no second sides. The sense in which horizon entropy is entanglement entropy is much more subtle for one-sided black holes. It involves entanglement between different parts of the near-horizon system.

As a one-sided  black hole evaporates the entanglement that accounts for its interior degrees of freedom  disappears, and is gradually replaced by entanglement with the outgoing Hawking radiation. A principle of ``transfer of entanglement" can be formulated. According to the argument of Almheiri, Marolf, Polchinski and Sully, it is when the transfer of entanglement is completed at the Page time, that a firewall  replaces the horizon.

Alternatives to firewalls may suffer contradictions which are similar to those of time travel. The firewall hypothesis would be similar to Hawking's chronology protection conjecture.

 \medskip
  \noindent
  \end{abstract}

  \end{titlepage}

  \starttext \baselineskip=17.63pt \setcounter{footnote}{0}

\tableofcontents

\sc
\section{Introduction}

Complementarity requires that the interior of a  black hole be represented by  same \dof \ that describe the exterior geometry of a black hole. Entanglement plays a crucial role relating the interior and exterior, but as time evolves entanglements shift from the near-horizon region to the evaporation products of the black hole. It seems likely that when this transfer of entanglement has been completed, the fundamental character of  interior geometry will have drastically changed.

The model that we will use for  \dof \ of a black hole \cite{Page:1993wv} is crude: a collection of localized radiation degrees of freedom representing the thermal atmosphere above the horizon, coupled to a stretched horizon composed of qubits that interact so as to insure the
 fast-scrambling property \cite{Hayden:2007cs} \cite{Sekino:2008he} \cite{Susskind:2011ap}  \cite{Lashkari:2011yi}.
Sub-Planckian physics on the near side of the horizon, i.e., the exterior region, is represented in a straightforward local way,  but the  physics behind the horizon is described by extremely subtle non-local entanglements.
In  ordinary systems  correlations of this type are not of much interest because they are far too complex and subtle---we will use the term fine-grained---to be detectable. For example they tend to involve correlations among huge numbers of \dof, which in practice are totally beyond practical observation. But to succinctly state the point of this paper\footnote{S. Shenker, Private communication.}:

\it What is  coarse-grained to an infalling observer, is extraordinarily fine-grained to an external observer.\rm

The super-fine-grained entanglements that account for the interior \dof \ degrade as the black hole ages, and completely disappear---more accurately they are transferred to the Hawking Radiation---by the Page time. The most obvious guess about what this means is that a firewall forms and replaces the usual smooth horizon.

The heart of this paper is in Sections 4 and 5. In Section 4 the importance of entanglement for the interior of the a black hole is explained. In Section 5, the disappearance of entanglement as the black hole evaporates is quantified.

\sc
\section{Three Ages of a Black Hole}

The lifetime of a black hole can be divided into three stages. During all three,  complementarity requires that  every meaningful question about the black hole can be formulated in terms of the quantum mechanics of a  single set of  \dof \  which describe the exterior of the black hole through conventional Hamiltonian quantum mechanics. The interior description of physics behind the horizon, when it makes sense, must be a reconstruction from those exterior degrees of freedom.
\subsection{Youth}

The first or $young$ stage occurs before the black hole is fully thermalized. This is the pre-scrambled phase which in Planck units\footnote{We use Planck units throughout. } lasts a time of order $$M \log{M}.$$  During this period, information behind the horizon is relatively easily expressed in terms of exterior degrees of freedom using the  \it \pb \ \rm procedure \cite{Freivogel:2004rd}, to be described shortly.

\subsection{Middle Age}

The \it middle-aged    \rm stage (the stage of decline) begins at the scrambling time $t\sim M \log{M},$ and lasts until the Page time \cite{Page:1993wv} at which half the entropy of the black hole has been radiated. In using  the ``pull-back---push-forward" procedure  after the scrambling time, one discovers that the pull-back cannot be done without encountering transplankian modes\footnote{I am indebted to Douglas Stanford for a discussion of this point.}. During early middle age,
information behind the horizon is reconstructed from entanglements between  different parts of the exterior of the  black hole. The \pb \ yields exterior operators that are scrambled over the stretched horizon, but not so delocalized as to involve the far-away Hawking radiation, at least at first.

During this second stage there is a gradual transfer of entanglement from the black hole to the Hawking radiation.

\subsection{Old Age}

\it Old age \rm  begins at the  Page time when the Hawking radiation has more entropy then the black hole. By that time the near-horizon region  is drained of entanglement entropy. What happens to the interior of the black hole during old age is not entirely  clear. The lack of entanglement in the near-horizon region means that there are no \dof \ in this region from which to reconstruct the interior. There appear to be two possibilities. The first, and in many ways the most straightforward,  advocated in \cite{Almheiri:2012rt}, is that a singular  firewall forms, destroying the interior of the black hole. The firewall at or slightly behind the horizon, and is not detectable by an infalling observer until the horizon is reached \cite{Susskind:2012rm}.

 The alternate viewpoint rests on the fact that black hole becomes entangled with Hawking radiation.
 This raises the possibility  that the \dof \ of  extremely distant quanta can be enlisted in reconstructing the interior region. If this is feasible it may avoid the need for firewalls, but there is grave danger of contradictions
occurring if information in the Hawking radiation is brought back into the black hole.

\subsection{Pull-Back---Push-Forward}

The assumption that everything about a black hole can be described in terms of exterior \dof \ does not preclude a description in the infalling frame. It does imply that the $interior$ \dof \ that  infalling observers use to organize their observations, can be constructed out of the exterior \dof.

\begin{figure}[h!]
\begin{center}
\includegraphics[scale=.3]{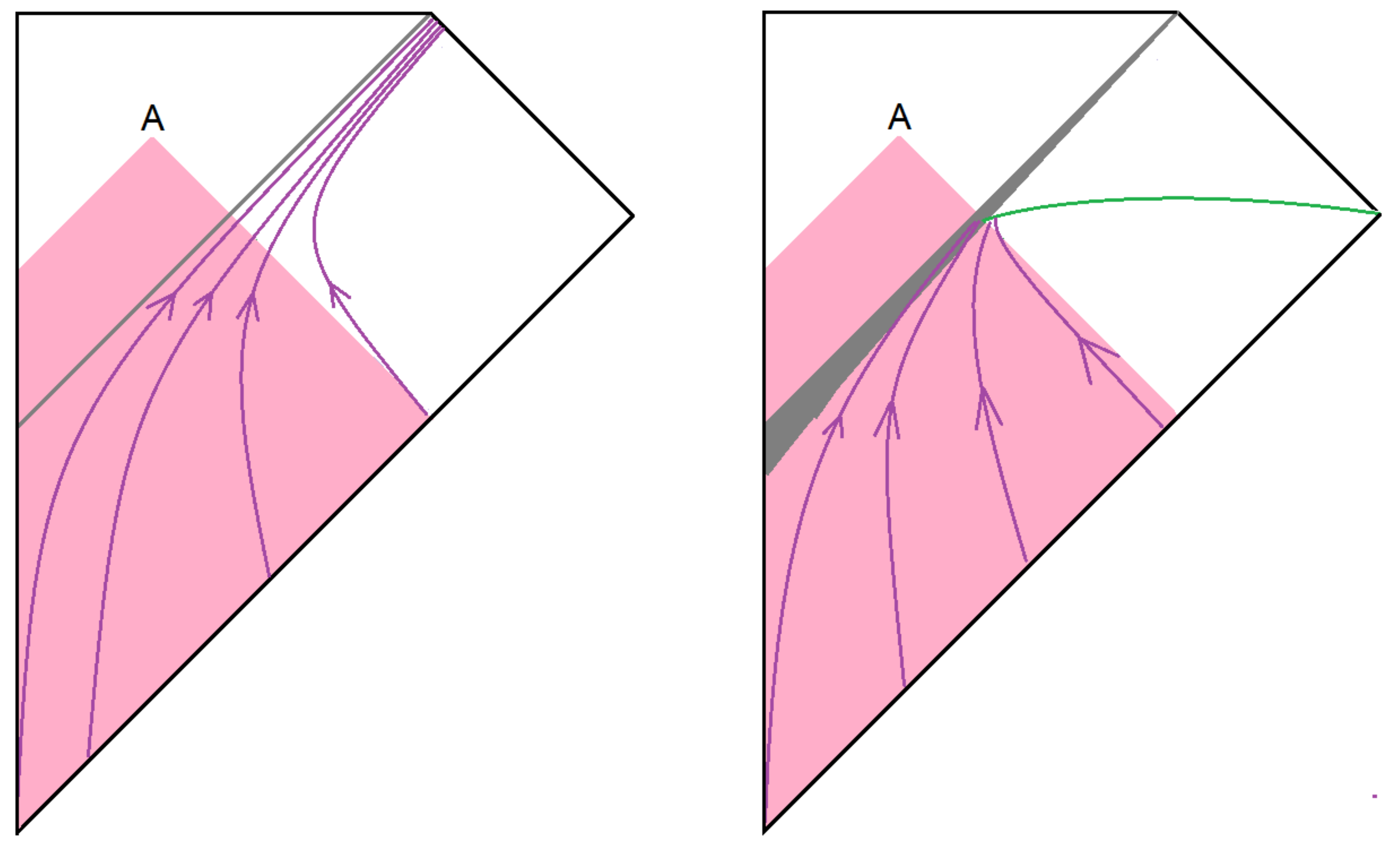}
\caption{The \pb \ strategy. The field operator is pulled back to the remote past using the  equations of motion
in the causal past of $A$ shown as pink. Then it is pushed forward using the exterior equations of motion. On the left the operator is pushed all the way to the infinite future in order to express it in terms of Hawking radiation. On the right the operator is pushed forward to a finite time-slice where it is expressed in terms of the exterior \dof.}
\label{pbpf}
\end{center}
\end{figure}

One proposal for identifying interior  with exterior \dof \ is the \pb \ strategy \cite{Freivogel:2004rd}. Consider some field operator $A_{int}$ behind the horizon, as in Figure \ref{pbpf}. The subscript $int$ indicates that it is an operator in the interior of the black hole.
The simplest version of \pb \ is to use the low energy equations of motion in the infalling frame to re-express $A$ in terms of an operator in the remote past. That operator is given by
\be
A_{past} = U^{\dagger}A_{int} U,
\ee
where $U$ is the unitary time evolution operator, in the infalling frame, that evolves from time $-\infty$ to the time of $A_{int}.$ Next push the operator forward to the remote future using the S-matrix in the exterior description.
\be
A_{future} =[ SU^{\dagger}] \ A_{int} \ [ U S^{\dagger} ].
\label{classic pbpf}
\ee
The $S$ matrix is the part of \pb \ in which the difficult physics of quantum gravity is buried.

This version of the \pb \ procedure expresses interior information in terms of the future asymptotic states of the Hawking quanta. It does not give a good picture of the evolution of that information with time. To address questions such as the interior of the black hole or the existence of firewalls \cite{Almheiri:2012rt}, we need a more local description, both in space and time.

\begin{figure}[h!]
\begin{center}
\includegraphics[scale=.3]{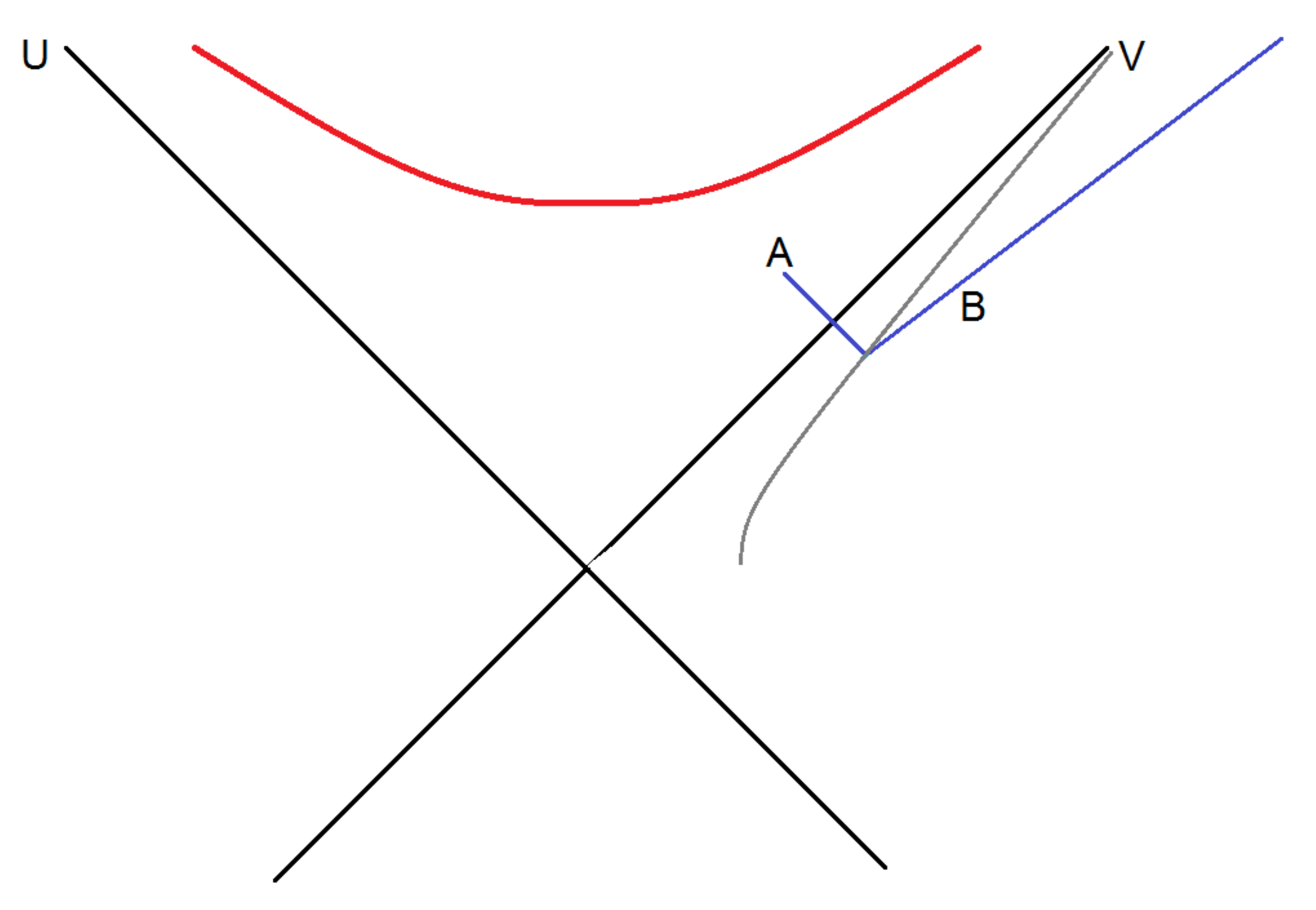}
\caption{A \S \ time can be identified with every point behind the horizon. The grey curve represents the stretched horizon about a Planck length outside the horizon. The modes $A$ and $B$ are on opposite sides of the horizon at the same time.}
\label{ta}
\end{center}
\end{figure}

The \pb \ procedure that we will consider  is more local in time. Instead of pushing all the way to the remote future, we can push forward to a finite \S \ time. To that end let us identify an exterior \S \ time for the operator $A.$  A convenient  way to do this is to begin with Kruskal coordinates $U,V$ as illustrated in Figure \ref{ta}

\begin{figure}[h!]
\begin{center}
\includegraphics[scale=.3]{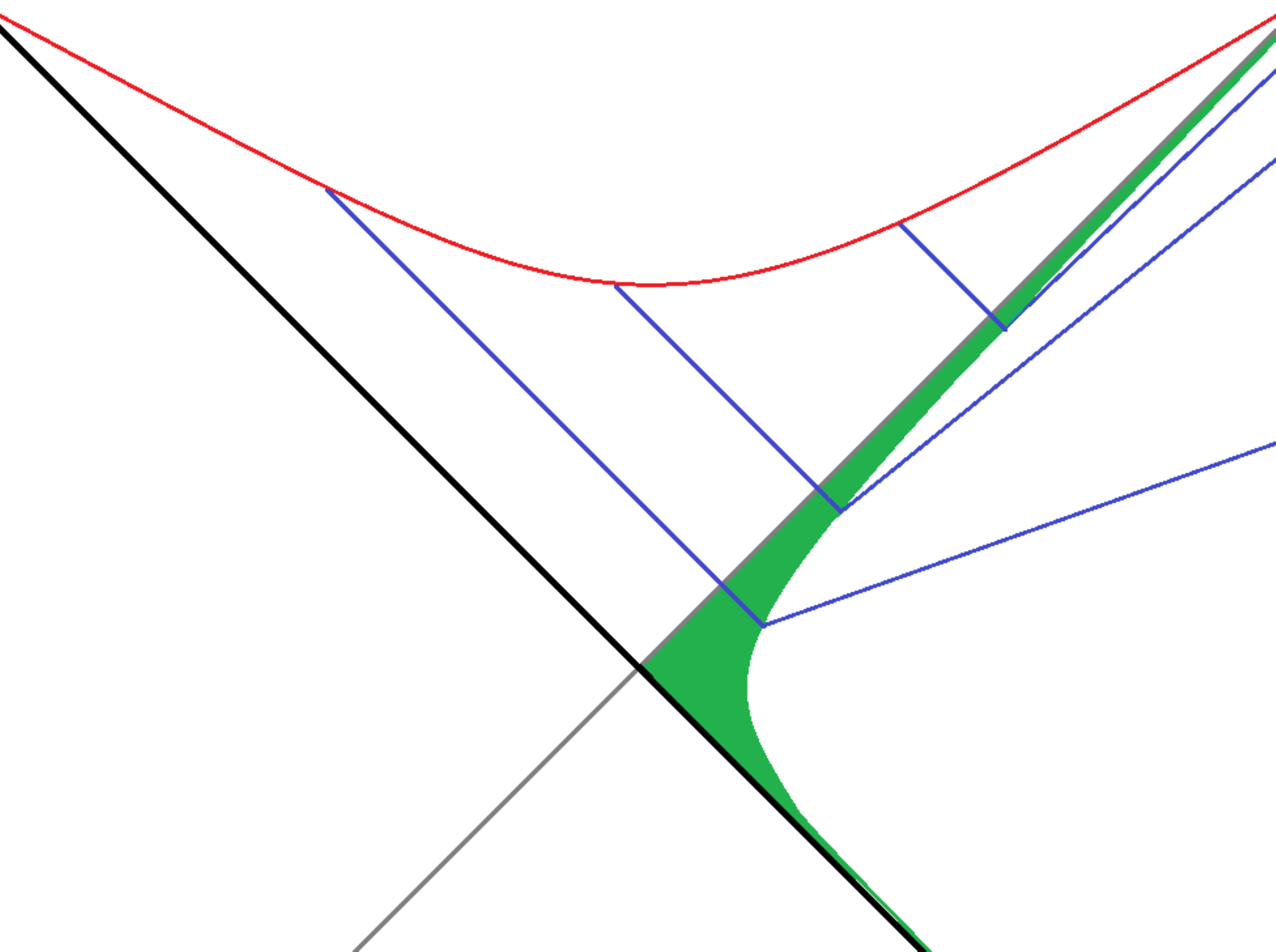}
\caption{Foliating the geometry by constant $t$ slices. Outside the stretched horizon the slices are defined by \S \ time. Inside the horizon they are light-sheets.}
\label{f9}
\end{center}
\end{figure}

 Each surface of constant $V$ intersects the stretched horizon\footnote{The stretched horizon will be identified as a surface about one Planck length outside horizon.} at some \S \ time given by
\be
t = R \log{\left( \frac{RV}{l_p} \right)}
\ee
If the operator $A$ is located at Kruskal coordinate $U_A,V_A$  an   exterior \S \ time will be associated with it:
\be
t_A =  R \log{\left( \frac{RV_A}{l_p} \right)}
\ee
Figure \ref{f9} shows the foliation of the black hole geometry by slices of constant $t.$

Let the time evolution operator in the exterior frame of reference be ${\cal{U}}(t_1, t_2).$  The generalization of \ref{classic pbpf} \ for finite time is
\be
A_{ext} = [ {\cal{U}}(-\infty, t_A) \ U^{\dagger} ] \ A_{int} \ [ U \ {\cal{U}}^{\dagger}(-\infty, t_A)]
\ee

During the young stage of evolution, the pull-back procedure is straightforward; no Planckian modes are encountered and the pull-back can be done using the low energy theory in the infalling frame. If the original operator  were outside the black hole, the \pb \ procedure would be trivial; push-forward would undo the pull-back and the result would be the original operator. But for an interior operator, a portion of the push-forward must use the Planckian stretched horizon degrees of freedom. For the young black hole, if $t_A$ is much less than the scrambling time, the push-forward will not severely delocalize the infalling information. In particular if  $U_A/V_A<<1$ (in other words the operator $A_{int} $ is close to the horizon) then the operator $A_{ext}(t_A)$ will be localized nearby $A_{int},$  with a small spread in the angular coordinates. In Planck units the angular spread will be of order \cite{Sekino:2008he}
$$ \frac{1}{M} \exp{\left( \frac{t_A}{4M}\right)}.$$  For $t_A<<M\log{M}$ this is small.

During early middle age when the time $t_A$ is later than the scrambling time, the push-forward process will involve a long enough time for the operator  $A_{ext}$ to be fully scrambled. It  will then become thoroughly delocalized over the entire stretched horizon.

Finally, we will see that by the Page time, $A_{ext}$ will no longer be found anywhere near the black hole. Instead it will be diffused throughout the first (early) half of the Hawking radiation. Or, as suggested in \cite{Almheiri:2012rt} \cite{Susskind:2012rm}, the whole construction loses its meaning, because by that time there is no longer a black hole interior.

\sc
\section{Entanglement and Scrambling}
\subsection{Maximal Entanglement}

The concepts of entanglement and scrambling of information will play a crucial role in understanding the relation between the
interior and exterior \dof \ of a black hole. We'll review them here.

We begin with the concept of maximal entanglement. Suppose $A$ and $B$ are two subsystems  that together comprise the system $(AB),$ and that $(AB)$ is in a pure state. The dimensions of the Hilbert spaces of $A$ and $B$  are $|A|$ and $|B|.$ Let us also assume that $|B| \geq |A|.$

Let $|a\rangle$ be an orthonormal basis for the Hilbert space of $A.$
The identity operator in the $A$ Hilbert space is given by
\be
I_A = \sum_a |a\rangle \langle a|
\ee

Any state of the combined system can be written in the form,
\be
\frac{1}{\sqrt |A|} \sum_a |a \rangle |a\}
\label{Psi}
\ee
where the curly bracket indicates a ket vector in the $B$ Hilbert space. In general the vectors $|a\}$ are nothing simple in terms of the \dof \  of $B.$ They are  defined by \ref{Psi}. In particular the $|a\}$ are not generally orthonormal. Normalization of \ref{Psi} requires
\be
\sum_a \{a|a\} =|A|.
\ee

In the special case in which
\be
\{a|a'\} = \delta_{aa'}
\ee
the system is said to be maximally entangled. The density matrix of the $A$ subsystem is proportional to the identity, and the entropy of both subsystems is given by $\log{A}.$ For the $A$ subsystem this is the maximum entropy possible.

Consider any observable in $A,$
\be
{\cal{O}}_A = \sum_{aa'}  O_{a a'} \ |a\rangle \langle a'|
\label{cor1}
\ee

A corresponding operator exists in the $B$ subsystem.

\be
{\cal{O}}_B = \sum_{aa'} O_{a a'} \ |a\} \{ a'|
\label{cor2}
\ee

The meaning of maximal entanglement is that for every observable in $A,$ one can predict the result of measuring it by measuring the corresponding observable in $B.$

In a maximally entangled state the density matrix of $A$ is proportional to the identity,
\be
\rho_A= \frac{1}{|A|} I_A.
\label{rho=I}
\ee
It follows that  the entanglement entropy is maximal,
\be
S = - \rho \log{\rho} = \log|A|.
\label{S=max}
\ee
In fact it is easy to prove that if \ref{rho=I} (or equivalently \ref{S=max}) is true, then the state is maximally entangled.

\subsection{Scrambling and Almost Maximal Entanglement}

Now let us consider the definition of a scrambled state \cite{Page:1993wv}  \cite{Hayden:2007cs} \cite{Sekino:2008he} \cite{Susskind:2011ap} \cite{Lashkari:2011yi}  for a system $C.$ A scrambled state is one such that if the system is subdivided into any two subsystems $(AB)$, the subsystems are almost maximally entangled. More exactly the entanglement entropy of the two subsystems is within a single bit of being maximal.  Moreover the density matrix of the smaller subsystem is very close to  the form in \ref{rho=I}.

Scrambled states can be generated by picking a random state  from the Haar-invariant ensemble of states of the composite system $C.$ If $A$ and $B$ are  equal size subsystems, the entanglement entropy is \cite{Page:1993wv}
\be
S_{ent} = \log{|A|} -\frac{1}{2}.
\ee
In other words the entanglement entropy is extremely close to maximal but not exactly. The slight deficit of $\frac{1}{2}$
implies that the correspondence implied by \ref{cor1} and \ref{cor2} is not quite faithful. In fact the mapping $|a\rangle \to |a\}$ is not unitary. The dimension of the  space spanned by the $|a\}$ vectors is smaller than $|A|$ by a factor $e^{-1/2} \approx 3/5.$ The implication is that there is some degree of imprecision---about a half of a bit of information---in predicting the results of measurements in $A$ by making corresponding measurements in $B.$ This imprecision decreases if we allow $|B|>|A|.$

Maximal entanglement is said to be monogamous. An example is a system of three spins. If two are entangled, for example in a singlet state, the third cannot be entangled at all. More generally, if a system in a pure state is divided into three subsystems and two are maximally entangled, the third must be in a pure state by itself. But when considering large scrambled systems it is possible to have overlapping almost-maximal entanglements. Consider a system of $2N$ qubits, $(q_1, q_2,...,q_{2N}).  $ Let us pick out $q_1$ for study. Assuming the system is scrambled, there are many decompositions of the remaining qubits that place $q_1$ in a subsystem $A$ which is $\leq N.$ For example $q_1$ belongs to the first half of the qubits, $(q_1, q_2, ...q_n).$ Or it belongs to the subsystem of odd qubits, $(q_1, q_3, q_5, ...).$ In each case $q_1$ is  close to being maximally entangled with the other subsystem, namely the second half or the even qubits.

A similar thing can be said about small sets of qubits. For example, we could choose to study the $(q_1,q_3)$ subsystem.  All of the operators composed out of both qubits are predictable by making a measurement in either the second half, or the even, qubits.

Consider two subsystems, each greater or equal to half the qubits,  which are almost non-overlapping. For example $(q_2, ....,q_{N+1})$  and $(q_{N+1}, ....,q_{2N}).$ Each of these subsystems are almost maximally entangled with $q_1,$ despite the fact that they only share one qubit. Thus there is a sense in which a subsystem can be almost maximally entangled with multiple almost non-overlapping subsystems.

\subsection{Maximal Entanglement and Bell Pairs}

Let us consider a system of $2N$ qubits grouped into two subsystems, $A$ and $B.$ A maximally entangled state can be constructed by pairing the qubits; one from $A$ and one from $B,$ into maximally entangled Bell states. The bell states can be chosen as singlet states,
\be
|S\rangle = \frac{|+,-\rangle -|-,+\rangle }{\sqrt 2}
\ee
The maximally entangled state $|E\rangle$ consists of $N$ such Bell states.
\be
|E\rangle = |S\rangle^{\bigotimes N}
\ee
There are many other maximally entangled states but they are all related to $|E\rangle$ by unitary transformations that act on $A$ and $B$ separately,
\be
|E'\rangle = U_A \ U_B  |E\rangle.
\ee
Each such state has maximal entanglement entropy
\be
S= N \log 2.
\ee
One can make less entangled states in an obvious way. Instead of $N$ Bell pairs, construct $M$ ($M<N$) bell pairs and arrange the remainder in a product state such as all qubits in the $|+\rangle$ state. This creates a state with entanglement entropy,
\be
S= M \log 2.
\ee
If this state is acted on by $U_A \ U_B$ the entanglement entropy remains equal to $M \log 2$ but the obvious pairing of the $M$ Bell pairs may be very obscured, particularly if the unitary matrices  $U_A , \  U_B$ are chosen randomly. Thus we see that entanglement entropy is a measure of the number of hidden Bell pairs that can be \it distilled \rm by applying unitary rotations to $A$ and $B.$

\subsection{Entanglement for Mixed States}

Entanglement entropy is a measure of how much you can learn about a system by observations on a second system, given that together they are in a pure state. For example, the entanglement entropy between a single qubit and the remaining $(2N-1)$ qubits in the previous scrambled example is exponentially close to $\log{2}.$ In a perfectly entangled state it is exactly $\log{2}.$ So we see that if the system is scrambled, by measuring  $(2N-1)$ qubits we can be determine everything possible about the first qubit, to within exponential accuracy.

If a system is divided into three subsystems, then typically no two will be in a pure state. Nevertheless, we may want to know how entangled two subsystems are. To give meaning to this question we need a measure of entanglement between $A$ and $B$ given that the combined system is in a mixed state $\rho_{AB}.$  There are a variety of entanglement measures which are generally very difficult to compute but fortunately interesting bounds can be placed on them. We will consider two of the better known measures. 

 The first is called
 \it relative entropy of entanglement \rm \ or $REE$ \cite{Plenio:2007zz}.

To define the $REE$ we first make some preliminary definitions. A separable state for a composite system $(AB)$ is defined to be any density matrix of the form
\be
\rho_{AB} = \sum_i P(i) \rho^i_A\bigotimes \rho^i_B
\label{separable}
\ee
where $P(i) $ is a normalized probability function. The correlations between $A$ and $B$ for a separable state are purely classical and do not involve quantum entanglement. Any good measure of entanglement that separates out classical correlation should give zero for a separable state \cite{Plenio:2007zz}.

To proceed,  define the relative entropy for two density matrices $\rho$ and $\sigma$
\be
S(\rho || \sigma) = Tr (\rho \log{\rho} - \rho \log{\sigma})
\label{RE}
\ee
This is a measure of how different $\rho$ is from some specific density matrix $\sigma.$ The relative entropy is greater or equal to zero.

The REE is defined by choosing $\sigma$ to be the separable state most resembling $\rho.$
\be
REE = \min_{\sigma} \frac{S(\rho || \sigma)}{2}
\label{REE}
\ee
where the minimization is over all separable states.
(The factor of $1/2$ is not conventional. It has been included so that $REE$ agrees with the entanglement entropy when $S_{AB} =0.$)

The $REE$ does a  good job of separating  quantum entanglement  from classical correlation but it is probably not exactly what we want. What we really want to know is how many entangled $A,B$
Bell pairs are hidden in the mixed state   $\rho.$
This brings us to the entanglement measure known as \it distillable entanglement,\rm  \ which we call $D.$
The reader is referred to \cite{Plenio:2007zz} for a precise definition of $D$. Here we will just remark that
the distillable entanglement is defined by a process of distillation, which in effect counts
 the number of hidden Bell pairs shared between $A$ and $B.$ This is what we need in order to bound the number of interior \dof \ that can be encoded in the exterior \dof \ of a black hole.

Both $REE$ and $D$ are well defined but very difficult to compute; however, easy to 
bound. In particular $D\leq REE.$ Furthermore $REE$ is easily bounded by a familiar quantity; namely the mutual information between $A$ and $B.$
To see the bound on $REE$  we substitute the separable state  $\rho_A\bigotimes\rho_B$ for  $\sigma$ in \ref{REE},

\be
S(\rho_{AB} || \rho_A\bigotimes\rho_B) = \frac{S_A +S_B -S_{AB}}{2}
\label{goo}
\ee
where $S_A,$ $S_B,$ and $S_{AB}$ are the entropies of  $A,$ $B,$ and $(AB).$  The quantity of the right side of \ref{goo} is the half the mutual information of $A$ and $B.$ Call it $\mu.$

From \ref{REE} and \ref{goo} we see that $ REE$ satisfies the inequality
\be
REE\leq \mu.
\ee

There are two important cases in which $D= REE = \mu.$ The first when the density matrix $\rho_{AB} $  is pure. In that case  $\mu$ and $REE$ are both equal to the usual entanglement entropy. The other case is when $\mu = 0.$ Since $D$ and $REE$ are positive or zero, and are also bounded by $\mu,$ they must vanish if $\mu$ vanishes.

Here are some examples of mutual information for a scrambled system.

\subsection{Examples of Mutual Information}

First consider a system of qubits,  divided into two subsystems  of $N$ and $M$ qubits, with $N\geq M.$ In this case the $\mu$ is just the entanglement entropy, which according to Page \cite{Page:1993wv} is given by
\be
\mu = S_{ent} = N \log{2} - 2^{M-N-1}.
\label{page formula}
\ee
For fixed $M$ and large $N$ this is exponentially close to the maximum, namely $N\log{2}.$ Thus one can essentially predict everything about the smaller subsystem from experiments on the larger subsystem.

For $N=M$ Page's formula reduces to
\be
\mu = N\log{2} - \frac{1}{2}
\label{half}
\ee

For large $N$ it seems that one can learn almost everything about one half of a scrambled system by measuring the other half, but the finite deficit $-\frac{1}{2}$ does represent a limit on the fidelity with which this can be done. To see how much of a limitation this is, consider the mutual information between half the system ($N $ qubits) and a single qubit in the other half.
\be
\mu =\frac{1}{2} (S_1 + S_N -S_{N+1})
\ee
In this formula $S_1$ is the entropy of the single qubit; $S_N$ is the entropy of the half not containing that qubit; and  $S_{N+1}$ is the entropy of the combined system containing the single qubit and the $N$ qubits. Using Page's formula \ref{page formula} gives
\be
\mu = \log 2 - \frac{1}{8}.
\ee
In this case the deficit is about $20$ percent of  the maximum, namely $\log{2}.$ By observing half the system one can do pretty well in predicting a single qubit in the other half, but there is a finite loss of fidelity.
Obviously  better fidelity can be achieved if one is allowed to look at all the other qubits, and not just half the total.

\subsection{Coarse-Grained Entropy}

The coarse grained entropy of a large thermodynamic quantum system  is defined in such a way as to be extensive. It ignores certain subtle correlations such as those that would distinguish a pure state from a mixed state.
It is constructed  by dividing the system into small cells. Each cell is bigger than some thermal correlation length but much smaller than the whole system. If the system is in its ground state then the entanglement entropy of each cell will be dominated by an area contribution, but if the system is at finite energy density this will not be so. In particular if the cells are big enough their entanglement entropy will be proportional to their volume. We assume this is the case.

 The entanglement entropy of the $n^{th}$ cell is $S_n.$ The coarse grained entropy is simply the sum of the $S_n.$
There is some ambiguity in this definition because of the surface effects in each cell, but if the theory is regulated and the cells are sufficiently large the  ambiguity is very small. The coarse-grained entropy ignores certain subtle non-local correlations  between subsystems.

 The quantum-information-theoretic model that has been applied to black hole physics envisions the black hole to be a collection of $N$ qubits in a scrambled  pure state \cite{Page:1993wv} \cite{Hayden:2007cs}.  The fine grained entropy of the whole system is zero but the coarse grained entropy can be computed by adding  the entanglement entropies of the individual qubits. To within exponentially small corrections,
 $$
 S=N \log{2}
$$
 In fact,  Page proves that the coarse grained entropy of a scrambled system is very well approximated (to within less than half a bit) by dividing the system into two equal subsystems and adding their entanglement entropies. In other words,  a small subsystem  can mean  anything up to  half the total system.

 The working hypothesis of the qubit model is that the Hawking Bekenstein entropy is the coarse-grained entropy of a scrambled  pure state.
In the scrambled   pure state model, the density matrix of a small subsystem  is proportional to the identity and therefore describes infinite temperature. The temperature of a black hole is not infinite, in fact large black holes are very cold, and so the approximation needs justification. A mathematical proof is far beyond the goal of this paper but some explanation is needed.

Let's regulate the system---for the moment not necessarily a black hole---by  dividing it into \it thermal cells \rm which are just big enough to contain about one bit of entropy. For a gas of radiation this means cells of size equal to the thermal wavelength. Also imagine truncating the space of states in each cell so that the energy is never much larger than the temperature. We will assume that this does not do much damage to the description of the system.

Now consider the Boltzmann distribution of each cell. It will have the thermal form,
\be
e^{-\frac{E}{T}}
\ee
For the range of allowed energies this is roughly constant. Thus for the regulated system, which by construction cannot have energy much greater than $T$ in each cell, the density matrix may be taken to be approximately random.

The model of thermal cells as qubits is clearly  very crude. The states that it describes are those that contain quanta of wavelength no shorter than the thermal wavelength, and at most one or two such quanta.

\setcounter{equation}{0}
\section{Black Holes and Entanglement Entropy}

Black hole entropy is entanglement entropy but the sense in which this is so is subtle. We begin with the not-so-subtle case of the so-called eternal black hole.

\subsection{Two-Sided Black Hole}

\begin{figure}[h!]
\begin{center}
\includegraphics[scale=.3]{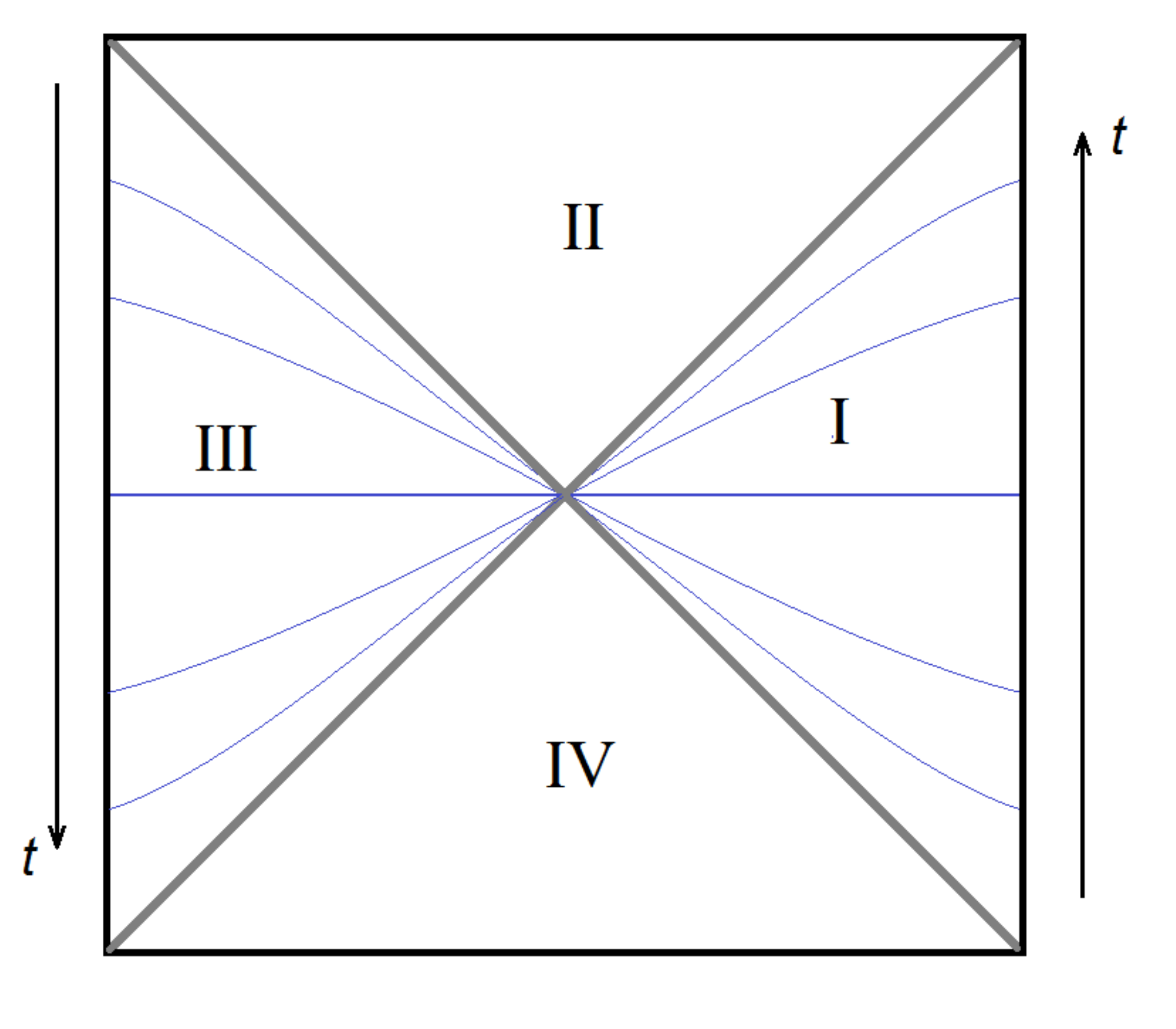}
\caption{Penrose Diagram for an ADS black hole. The diagram is composed of four quadrants. Quadrant I and II represent the physical black hole.}
\label{f1}
\end{center}
\end{figure}

It is often claimed that the entropy of a  black hole is due to the entanglement between the two sides of the horizon. This is correct both for Rindler space  and for eternal black holes, but is not correct for real black holes.  As an example of when it is correct,  consider the large ADS eternal black hole shown  in Figure \ref{f1}.
The quantum state of the system is described in terms of two copies \cite{Israel:1976ur} \cite{Maldacena:2001kr} \cite{Czech:2012be}  of a boundary conformal field  theory, the left and right copies. The two copies have identical  Hamiltonians except for sign. The Hamiltonian of the right copy is positive, and that of the left copy  is
negative. The total energies of the left and right sides add up to zero. Let the energy eigenvectors and eigenvalues of the right side be
$$|i\rangle_r$$ and $$E_i.$$ The corresponding quantities for the left side are,
$$|i\rangle_l$$ and $$-E_i.$$ The eternal black hole is described by the entangled state
\be
\Psi = \sum_i e^{-\frac{\beta}{2} E_i}|i\rangle_l|i\rangle_r
\label{thermofield}
\ee

By tracing over the left side one finds the density matrix for the right side to be the thermal density matrix,
\be
\rho = \sum_i e^{-\beta E_i}|i\rangle \langle i |
\ee
where for notational simplicity the subscript $r$ was left out.
The state $|\Psi\rangle$ is pure and has zero entropy, but the density matrix $\rho$ has entanglement entropy given by the Hawking Bekenstein value.

\begin{figure}[h!!]
\begin{center}
\includegraphics[scale=.3]{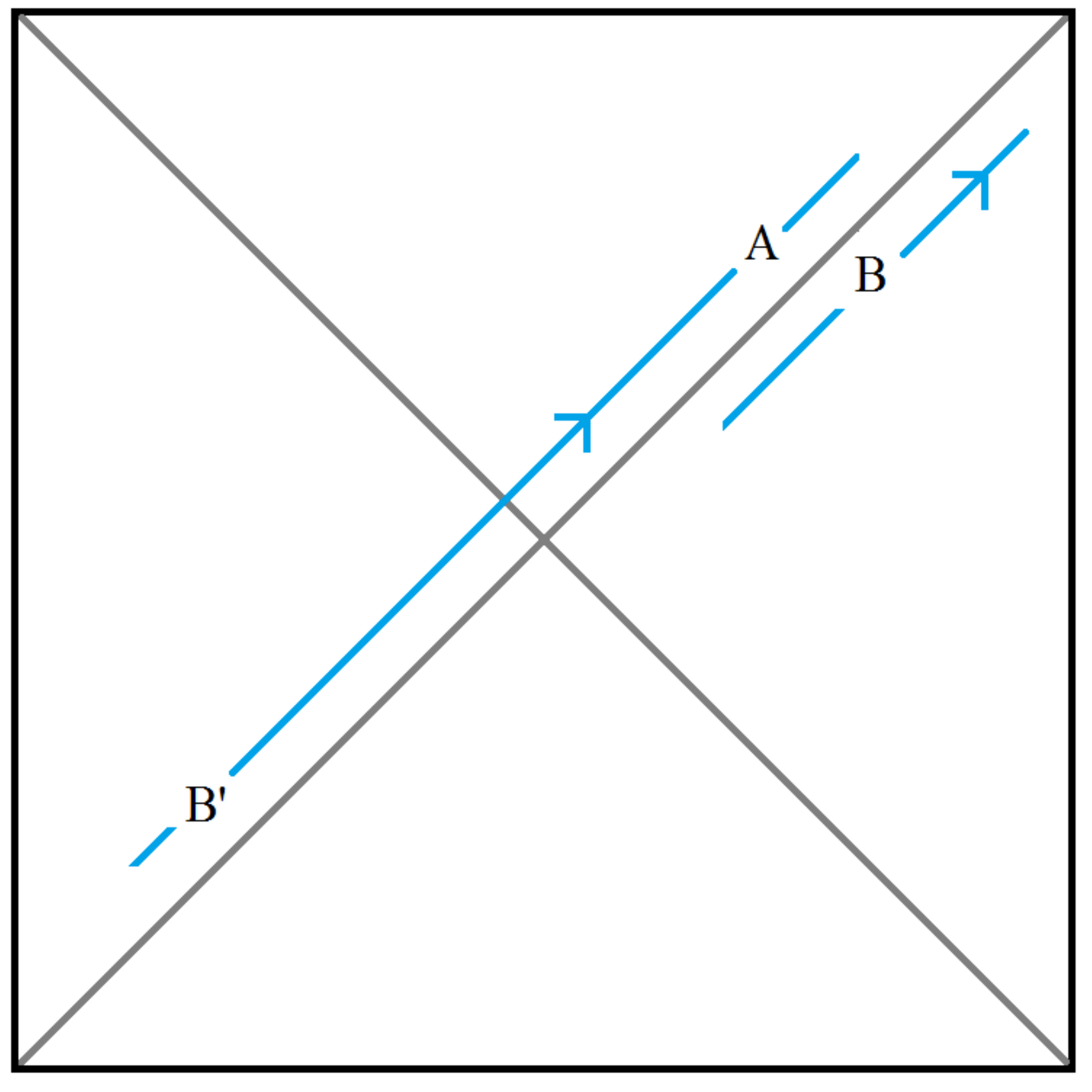}
\caption{The mode at $B$ is entangled with a mode at $B'$. Both  $B$ and $B'$ are outside their respective horizons. It is clear from the diagram that the mode $A,$ which is situated near $B,$ but behind the horizon, is the same as $B'.$ The strong entanglement between $A$ and $B$ is nothing but the entanglement between $B$ and $B'.$ }
\label{f11}
\end{center}
\end{figure}

If we think of the physical black hole being the right and upper quadrants I and II, then it is evident that the degrees of freedom behind the horizon, in quadrant II of the diagram,  partly evolve from degrees of freedom in quadrant III. This is illustrated in Figure \ref{f11}. What is clear from the diagram is that the degrees of freedom, perceived by an observer falling in from the right as being just behind the horizon in quadrant II, are in fact to be identified  with the degrees of freedom that generate the outgoing Hawking photons on the left\footnote{I thank Don Marolf for a helpful discussion on this point.}.

The precise statement of this duality is complicated by the fact that higher angular momentum modes and massive modes do not travel on radial null rays. A general mode at $A$ will trace  trace back partly to region I, and partly to region III. The contributions from region III are what are relevant when claiming that
 the entanglement between the left and right quadrants is important to reconstructing the  interior of the eternal black hole \cite{Czech:2012be}.

In the qubit model of the black hole, the thermally entangled state \ref{thermofield} is replaced by a maximally entangled state. Every operator on one side has a corresponding partner on the other side whose measurement (if it were possible) would predict the original operator.

\subsection{One-Sided Black Hole}

Real black holes, formed in collapse, don't have two sides\footnote{Louko, Mann, and Marolf \cite{Louko:2004ej} have considered a very interesting class of black holes called geons which have only one side that are obtained by an identification procedure from the two-sided case. Although these black holes cannot be created by collapse they may offer some insight into one-sided black holes.}.

\begin{figure}[h!!]
\begin{center}
\includegraphics[scale=.3]{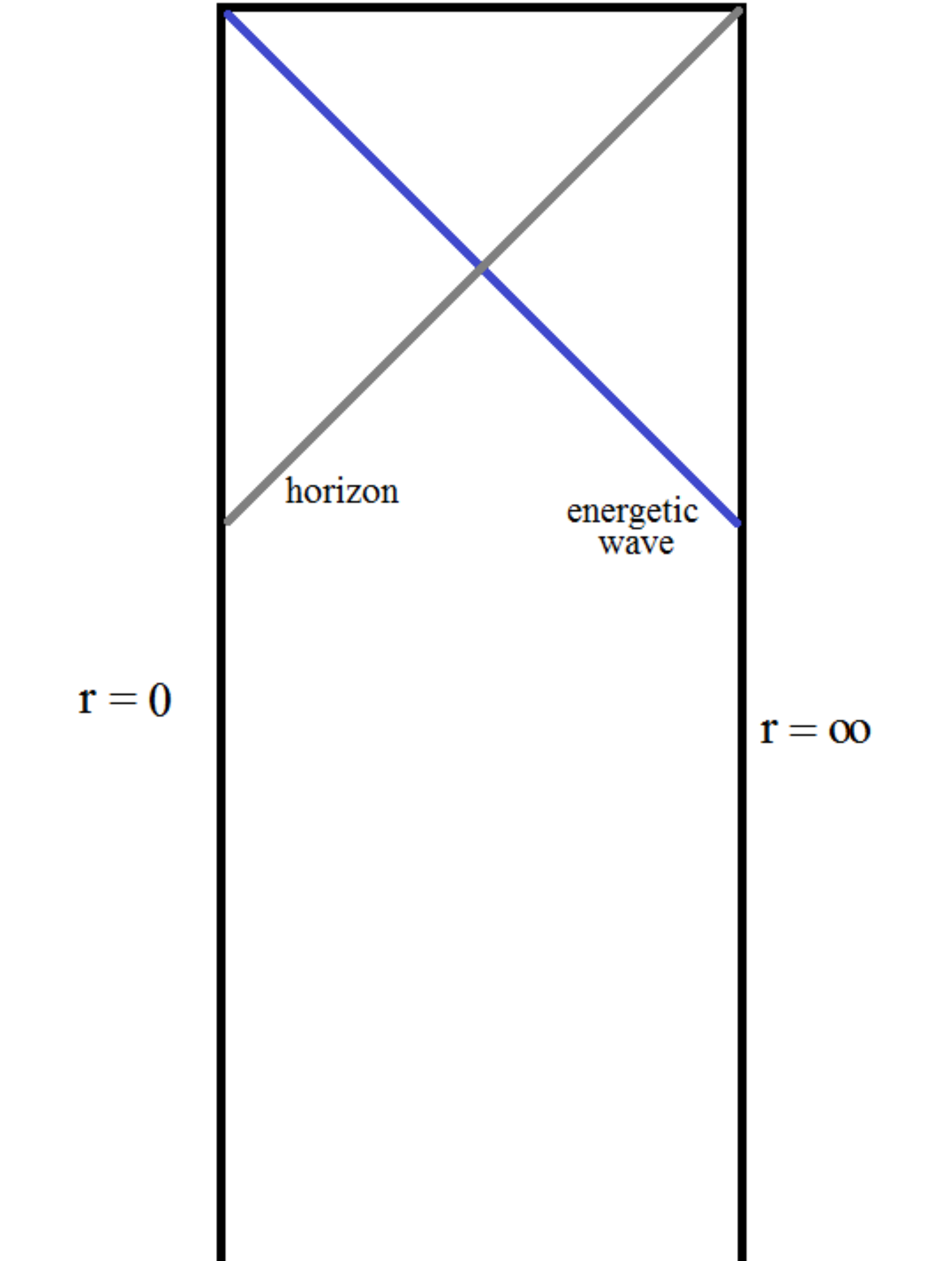}
\caption{A one-sided black hole formed by shooting in a wave from $r=\infty.$}
\label{f111}
\end{center}
\end{figure}

 In the ADS case where evaporation is not an issue, a black hole can be formed by shooting in a wave from a single boundary. The resulting black hole is described by a pure state of a single copy of the boundary theory as in Figure \ref{f111}.
 It
 therefore  has zero fine-grained entropy. The question is, what is the meaning of the Hawking Bekenstein entropy in this case? Is it entanglement entropy? Can it be formulated in terms of exterior \dof?
  The  obvious answer is that it is  coarse grained entropy. For a system in a pure state, coarse grained entropy is entanglement entropy, but not entanglement with a second system. Its origin is the entanglement of different parts of the same system.

One idea might be to divide the near-horizon region geometrically,   into two hemispheres---``north" and ``south"---and the coarse grained entropy can be identified with,
\be
S = S_N + S_S.
\ee
The $S_{N,S}$ would be entanglement entropies, which for an overall pure state satisfy $S_N = S_S = S_{ent}$ where $S_{ent} $ is the entropy of entanglement of the two hemispheres. The total entropy would be twice the entanglement entropy,
\be
S=2S_{ent}.
\ee

However, there are reasons to think that a division based on localizable \dof \ is not the right idea. In particular, the fast-scrambling property of black holes indicates that the stretched horizon dynamics is highly non-local, and best described by delocalized matrix degrees of freedom \cite{Sekino:2008he} \cite{Susskind:2011ap}. We will discuss a better decomposition of the black hole below.

\subsection{The Schwarzschild Black Hole}

A real Schwarzschild black hole, formed in collapse, is one-sided. Unlike ADS black holes, Schwarzschild black holes do not at present have a precise mathematical exterior quantum description. We will assume that such a description exists and that it involves degrees of freedom accessible to an outside observer.

For now we will make use of a simple description of the exterior
 \dof  \ based on dividing space into three regions. The metric has the usual form,
\be
ds^2 = -(1-\frac{R}{r})dt^2 + \frac{1}{(1-\frac{R}{r})}dr^2 + r^2 d\Omega_2^2
\ee

\bi

\item The distant region $r> \frac{3}{2}R$ beyond the centrifugal barrier will be called ${\cal{R}}.$ ${\cal{R}}$ stands for radiation. It contains whatever Hawking radiation may have been emitted. The degrees of freedom in this region are low energy effective fields including gravitons.  ${\cal{R}}$ will not be thought of as part of the black hole.

\item The \it zone\rm \footnote{The term zone in this context was coined by Raphael Bousso. Private communication.} ${\cal{Z}}$ is the near-horizon region extending from the stretched horizon to the centrifugal barrier at  $r\sim \frac{3}{2}R.$
    It is the region occupied by an Unruh thermal atmosphere that can be described by sub-Planckian field degrees of freedom, The zone is in thermal equilibrium at the local Rindler temperature $1/{(2\pi \rho)}$ where $\rho$ is the proper distance from the horizon. The region \z \ carries some small fraction of the total entropy of the black hole. If we nominally place the separation between Planckian and Sub-Planckian at $.1$ times the Planck energy, then the thermal atmosphere would have about one percent of the black hole entropy.

\item The Planckian stretched horizon    \h, is where most of the entropy resides. The \dof \ in  ${\cal{H}}$ are not organized into a local field theory on a two dimensional membrane. In order to have the fast-scrambling property, they must interact in a highly non-local manner. A system of $N\times N$ matrix degrees of freedom is the natural candidate for the \dof \ of ${\cal{H}}.$ The minimal value of $N^2$ that can capture the physics of a black hole of entropy $S$ would be $N^2 = S.$

\ei

Although it is not essential, we will assume that the \dof \ of \z \ and \h \ can be replaced by qubits. Each thermal cell in the zone may be represented by a single qubit since by definition it has about one unit of entropy. The same can be said of each of the $N^2$ matrix degrees of freedom in \h.

\subsection{Interior Degrees of Freedom and Entanglement}

We will assume that once the scrambling time has passed, the qubit degrees of freedom of \z \ and \h \ have been scrambled. This means, among other things, that \z \ is maximally entangled with \h.
We will explore
the  hypothesis that that the interior \dof \ of a black hole are described by the nonlocal entanglements between the degrees of freedom comprising the zone \z \ and the stretched horizon \h. Let's first consider the number of \dof \ that are required to describe the interior at a given time. Figure \ref{f9} shows the black hole foliated by constant time slices. The portion of each time slice inside the horizon is a light-sheet. One can assign a maximum entropy to such a light sheet \cite{Fischler:1998st} and in this case it is  the area of the horizon. With the usual identification between maximum entropy and  number of degrees of freedom, we find that the number of \dof \ of the interior is the same as the number describing the exterior; namely the Bekenstein Hawking entropy.

From the perspective of an infalling observer, the interior and exterior degrees of freedom are entangled.
As an example consider a pair of field modes $A$ and  $B$  located  on either side of  the horizon \cite{Almheiri:2012rt}. The field mode $B$ is in \z. It's partner $A$ is behind the horizon, but at a similar proper distance (time-like in the interior) from the horizon.

In the infalling frame $A$ and $B$ are entangled. The fundamental assumption of this paper is that the relation between $A$ and $B$ is paralleled by an equivalent relation between $B,$ and some subsystem $B'$ in the stretched horizon, ie., in the non-local matrix system. The subsystem $B'$ is not to be thought of as being entangled with $A.$ Rather
\it it is $A.$ \rm
\be
A=B'
\label{A=B'}
\ee

After the scrambling time but before any appreciable amount of Hawking radiation has been emitted, we expect that the  \dof \ in \z  \ will be almost maximally entangled with those in the larger system \h, as discussed in Section 3.2. Following the argument that led to \ref{page formula}, there is a subsystem of \h \ which is maximally entangled with the degree of freedom $B$ with very high fidelity. That subsystem is the exterior representation of the interior mode $A.$

We may want to be more ambitious and consider not just a single $A,B$ pair, but several simultaneously. In the infalling frame any system of modes $B$ will be maximally entangled with a corresponding system $A$ behind the horizon. In the exterior frame the subsystem of $B$ modes will be entangled with a subsystem $B'$ in the stretched horizon \h.

It is evident that the number of interior degrees of freedom that can be constructed from the exterior degrees of freedom is bounded by the number of entangled pairs shared between \z \ and \h.

\sc
\section{Decline}

We come now to the main point of the paper, which is to be as quantitative as possible about how  entanglements between \z \ and \h \ disappear as  the black hole evaporates.
The model we will use is by now familiar \cite{Hayden:2007cs}. Hawking radiation is described by transferring qubits from the black hole to the system of Hawking radiation at a steady rate. It is also assumed that the near-horizon system is kept scrambled by internal dynamics.

Since the black hole will become entangled with the Hawking radiation, near-horizon \dof \ will not remain in a pure state.
The degree of entanglement between \z \ and \h \ must be measured by the distillable entanglement.

Therefore the model will be supplemented with two  additional assumptions:

\bi
\item The interior degrees of freedom are constructed from entanglements between \z \ and \h. The \dof \  in  ${\cal{R}}$ (the Hawking radiation)  are not involved in the reconstruction of the interior.
\item The number  of interior \dof \  that can be encoded in the exterior \dof \ of  \h \
   is equal to the distillable entanglement, $D$ of the (\z,\h) system. It follows that the number of interior \dof \ is  bounded by half the corresponding
mutual information.
\ei

We consider a three component system of qubits consisting of \z, \h, and  the Hawking radiation \r. The number of qubits in each subsystem is $n,$ $N,$ and $R$ respectively. The ratio  $n/N$ (the number of \dof \ in the zone to the number in the stretched horizon) is some number less than $1,$ which depends on exactly how we divide the black hole into the subsystems \z \ and \h.

The  Bekenstein Hawking entropy of the black hole is
\be
S_{BH}= (N+n)\log{2}
\ee

The mutual information between \w \ and \W' \  is given by
\be
2\mu = S_{Z} + S_{H} - S_{ZH}
\ee

Because the entire system including the Hawking radiation is in a pure state, the entropy of the \z,\h \ system is equal to the entropy of \r. Thus
\be
2\mu = S_{Z} + S_{H} - S_{R}
\ee

At the early stages before the Page time, the biggest system is \h. This allows us to compute the various entropies using \ref{page formula}. The exponentially small deficits $2^{M-N-1} $ in \ref{page formula} are generally unimportant but we will keep them for completeness.
\bea
S_{Z} \eq n \log{2}  -2^{n-R-N-1}         \cr \cr
S_{H}\eq S_{R Z}=(R+n)\log{2} - 2^{R-N+n-1} \cr \cr
S_{R} \eq R\log{2} - 2^{R-N-n-1}
\label{S-case 1}
\eea
One finds,
\be
\mu = n\log{2} - \left(  2^{n-2} - 2^{-n-2} \right)   2^{R-N} -2^{n-R-N-2}
\label{case 1 mu}
\ee

Notice that for $R<<N$ this is exponentially close to the maximum entropy for an $n$ qubit system. It is also the number of \dof \ one expects in the interior region to be entangled with the zone \dof. But as the radiation increases, $\mu$ decreases, indicating  that the number of interior \dof \ that can be encoded  in \h \  \dof, decreases as the black hole radiates. That in itself is hardly surprising since the black hole shrinks. But the decrease in $\mu$ is faster than that. In fact \ref{case 1 mu} goes to zero when $R\approx N-2n.$ In terms of entropy this depletion of $\mu$ occurs at
\be
S_R = S_{BH} -3n
\label{zip}
\ee

However, this is  misleading. The middle equation of \ref{S-case 1} breaks down at the point where $N$ ceases to be greater than half the total number of qubits. The transition point is
\be
R = N-n.
\label{transpoint}
\ee

 At the transition point  \ref{transpoint} the $\mu$ is
\be
n\log{2} - \frac{1}{4}\left(   1-\frac{1}{2^n}              \right)
\ee
In other words it has barely changed from the maximum value $n\log 2.$

For $N< R+n$ the second of equations \ref{S-case 1} becomes,
\be
S_{H}= N\log{2} - 2^{N-R-n-1}
\label{S-case 2}
\ee
In this range $\mu$ is given by,
\be
\mu=\frac{n+N-R}{2}\log{2}+ 2^{R-N-n-2}-2^{N-R-n-2} -2^{n-R-N-2}
\ee

In terms of the black hole and radiation entropies $S_{BH} = (N+n)\log{2},$ $S_R= R\log 2,$ the mutual information is
\bea
\mu \eq \frac{S_{BH}- S_R}{2} + 2^{R-N-n-2}-2^{N-R-n-2} -2^{n-R-N-2} \cr \cr
&\sim&\frac{ S_{BH}- S_R}{2}
\label{linear}
\eea

At the Page time $S_{BH} = S_R,$ we find,
\be
\mu \approx \frac{1}{2}
\ee
Beyond the Page time, the mutual information vanishes.
 Since the number \dof \ behind the horizon is bounded by the mutual information, we conclude that
 the interior degrees of freedom have been completely depleted by the Page time.

 The interior remains depleted until the black hole has completely evaporated, or until new matter is thrown in to rejuvenate the horizon by creating new entanglements \cite{Susskind:2012rm}.

The presence or lack of entanglement between the zone and the stretched horizon is a property of the black hole state that in principle  has measurable implications for an observations of the near-horizon region. Among other things, one could determine by such observations  whether the black hole is young, middle-aged, or old.


\section{ Firewalls}
\subsection{From Special to Typical}

Let's review the evolution of a black hole from its early youth to old age. The black hole is formed by collapse in one of many ``special" states \cite{Susskind:2012rm} with smooth non-singular horizons. The special states are a small subset of black hole states. According to one estimate \cite{'tHooft:1993gx}, the logarithm of number of matter states that can collapse to a black hole is of order
$M^{\frac{3}{2}}$
which is smaller than the black hole entropy $\sim M^2.$ Although these states are macroscopically similar to one another, they are microscopically different. By the scrambling  time $M\log{M}$ these differences have been scrambled but they have  not been lost. In fact they are present in the form of the subtle entanglements between parts of the near-horizon degrees of freedom and the stretched horizon. These differences are in principle observable\footnote{This has also been pointed out by B. Freivogel and R. Bousso. Private communication.} to an observer falling through the horizon, but they are probably very dilute; the number of bits per unit area in these differences is of order
\be
\frac{M^{3/2}}{M^2} = \frac{1}{\sqrt{M}}
\ee
and goes to zero with increasing mass.

As the evolution proceeds through middle-age, the entanglements between parts of the stretched horizon and the zone begin to disappear, or more precisely, they are transferred to the Hawking radiation. The differences between the  special states are also transferred to the Hawking radiation. As  the Page time is approached the density matrix of the near-horizon region tends to a universal maximally incoherent form. An infalling observer with no access to the Hawking radiation cannot distinguish the difference between the various initial special states. More importantly there are no \dof \ remaining in the interior region.
An obvious guess is that a  firewall forms at the Page time \cite{Almheiri:2012rt}. As explained in \cite{Almheiri:2012rt},  the overwhelming majority of black hole states must have firewalls. Only the small fraction which are special don't, but these include all the states  that recently  formed from the standard process of collapse.
\begin{figure}[h!!]
\begin{center}
\includegraphics[scale=.3]{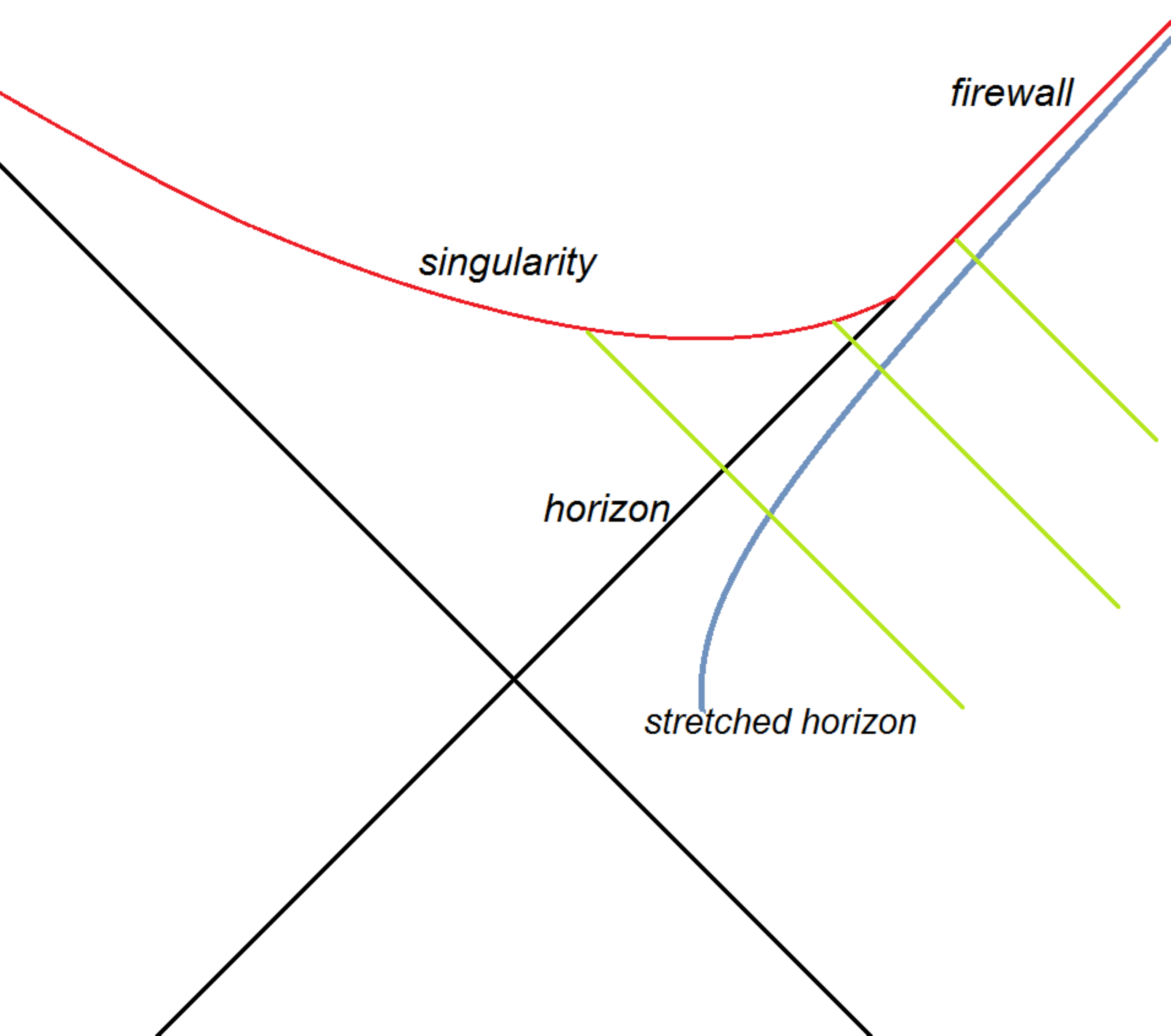}
\caption{A model of firewall formation in which the black hole singularity migrates from it classical location.}
\label{f5}
\end{center}
\end{figure}

In \cite{Susskind:2012rm} it was suggested that the depletion of the interior \dof \ can be described by the singularity gradually migrating away from its classical location, until it hits the horizon at the Page time, and turns it into a firewall. This was illustrated in Figure 4 of \cite{Susskind:2012rm} which is reproduced here as Figure \ref{f5}.
The squeezing out of the space between the singularity and the horizon is a pictorial representation of the decrease of distillable entanglement as the Page time is approached. It is possible that the the migration of the singularity may be too classical a description of the evolution of the interior space. Perhaps the interior degrades in some way that cannot be described by classical geometry.

\bigskip

\bigskip
\bigskip

\bigskip
One point that bears emphasis is that the mutual information is not generally a good measure of entanglement unless it is either close to maximal or close to zero. In general it is an upper bound on the distillable entanglement. It therefore underestimates the rate at which the interior geometry degrades. However, it seems reasonable to assume that some entanglement between \z \ and \h \ persists until the mutual information goes to zero at the Page time.

\subsection{Alternative to Firewalls?}

Is there an alternative to the firewall conclusion \cite{Bousso:2012as}\cite{Harlow:2012me}\cite{Nomura:2012sw}\cite{Mathur:2012jk}\cite{Chowdhury:2012tr}\cite{Banks:2012nn}? Perhaps.
The idea goes back to a point made in Section 3.4:

 \it Entanglements in the infalling frame between $A$ and $B$ are equivalent to entanglements between $B'$ and $B$ in the exterior frame. \rm

The case for firewalls is based on the assumption that $B'$ is a subsystem of the stretched horizon. However, from a purely information theoretic standpoint, the information in the Hawking radiation is also available for reconstructing $A.$  After the Page time, there is a subsystem, $R_B,$ of the Hawking radiation which is entangled with $B.$ We may hope to elminate the need for firewalls by replacing $B'$ with $R_B.$ In other words we replace \ref{A=B'} with,
 \be
   A = R_B.
   \ee
It is certainly true that information in quantum gravity can be delocalized in surprising ways. Whether utilizing \dof \  that are nominally extremely far away  is too high a price to pay, in order to avoid
         firewalls, remains to be seen, but there are two arguments that suggest that it is.

         Consider an infalling particle whose trajectory is aimed toward $A.$ In the infalling frame it will interact with $A,$ influencing and being influenced by it. How is that interaction represented in the exterior description?

         What one knows is that the infalling particle will affect the stretched horizon by classically perturbing it. Moreover, the microscopic information carried by the particle will diffused by the fast-scrambling process and will modify the entanglements present on the stretched horizon. Thus if one identifies $A$ with $B'$ there is opportunity for the infalling particle to interact, both in the infalling and exterior frames.

         But if $A$ is diffused over the Hawking radiation at a distance of order $M^3$ from the horizon, it is hard to see from the exterior point of view how it can interact with an infalling particle.

 Another danger was  implicit in \cite{Almheiri:2012rt} and was emphasized by Bousso and others\footnote{R. Bousso, private communication.}. To see what's at stake, consider the radiation-black-hole system at some late time, well past the Page time---for definiteness let's say that nine-tenths of the entropy has been radiated. In the infalling frame the mode $A$ is localized behind the horizon,   On the other hand the subsystem $R_B$ is in the first half of the Hawking radiation. Moreover, according to the above hypothesis, the information in $A$ is identical to that in $R_B.$

      Now suppose Alice is capable of extracting the information in $R_B$ from the early Hawking radiation. She can then throw it back into the black hole where it meets its alter ego, $A.$
      This has a distinct similarity with time travel. The late bit  $A$ has somehow been transported to the early Hawking radiation, and then brought back to the black hole, where it meets itself.

      There are two ways that nature could provide chronology protection  \cite{Hawking:1991pk} in this context. It may simply be physically impossible to distill $R_B$ out of the Hawking radiation in time to bring it back to meet $A.$ However, the ``mining" of ADS black holes \cite{Almheiri:2012rt} provides an example in which Alice has as much time as she needs, since the ADS black hole does not evaporate.

      The other possibility is that
 firewalls play the role of Hawking's chronology protection enforcer.

\section*{Acknowledgements}

I am very  indebted  to   Douglas Stanford, for a number of key insights. I am also grateful to Raphael Bousso, Ben Freivogel, Dan Harlow, Don Marolf, Joe Polchinski, Steve Shenker, and Mark Van Raamsdonk for helpful discussions.

\end{document}